\documentclass[a4paper,12pt]{article}
\newcommand{\be}{\begin{equation}}
\newcommand{\ee}{\end{equation}}
\newcommand{\bea}{\begin{eqnarray}}
\newcommand{\nn}{\nonumber}
\newcommand{\eea}{\end{eqnarray}}

\begin{document}

\begin{titlepage}

\title{\bf {Bouncing Cosmologies and Self-Gravitational
Corrections }}
 \author{M.R. Setare  \footnote{E-mail: rezakord@ipm.ir}
  \\{Physics Dept. Inst. for Studies in Theo. Physics and
Mathematics(IPM)}\\
{P. O. Box 19395-5531, Tehran, IRAN }}
\date{\small{}}

\maketitle
\begin{abstract}
A four-dimensional timelike brane is considered as the boundary
of the $SAdS_{5}$ bulk background. Exploiting the
CFT/FRW-cosmology relation, we derive the self-gravitational
corrections to the first Friedmann-like equation which is the
equation of the brane motion. The additional term that arises due
to the semiclassical analysis, can be viewed as stiff matter
where the self-gravitational corrections act as the source for
it. This result is contrary to standard analysis that regards the
charge of $SAdS_{5}$ bulk black hole as the source for stiff
matter. A very interesting feature of the solutions of FRW
equation is that they all (for $k=0,\pm1$) have a non-vanishing
minimum value for the scale factor.
 \end{abstract}
\end{titlepage}
\newpage

\section{Introduction}
Black hole thermodynamic quantities depend on the Hawking
temperature via the usual thermodynamic relations. The Hawking
temperature undergoes corrections from many sources:the quantum
corrections, the self-gravitational corrections, and the
corrections due to the generalized uncertainty principle.\\
 There
has been much recent interest in calculating the quantum
corrections to $S_{BH}$ (the Bekenestein-Hawking entropy)
\cite{maj1x}-\cite{set6}. The leading-order correction is
proportional to $\ln{S_{BH}}$. There are, {\it two} distinct and
separable sources for this logarithmic correction
 \cite{gg2x,maj3x} (see also recent paper by Gour and Medved \cite{ medved}).
  Firstly, there should be a correction
 to the number of microstates that is a quantum correction to the
 microcanonical entropy, secondly, as any black hole will typically exchange heat or
 matter with its surrounding, there should also be a correction due to thermal
 fluctuations in the horizon area. In a recent work Carlip
 \cite{carx} has deduced the leading order quantum correction to
 the classical Cardy formula. The Cardy formula follows from a
 saddle-point approximation of the partition function for a
 two-dimensional conformal field theory. This leads to the
 theory's density of states, which is related to the partition
 function by way of a Fourier transform \cite{carli}. \\
Concerning the quantum process called Hawking effect
\cite{hawking1} much work has been done using a fixed background
during the emission process. The idea of Keski-Vakkuri, Kraus and
Wilczek (KKW) \cite{KKW} is to view the black hole background as
dynamical by treating the Hawking radiation as a tunnelling
process. The energy conservation is the key to this description.
The total (ADM) mass is kept fixed while the mass of the black
hole under consideration decreases due to the emitted radiation.
The effect of this modification gives rise to additional terms in
the formulae concerning the known results for black holes
\cite{corrections}; a nonthermal partner to the thermal spectrum
of the Hawking radiation shows up.\\
The generalized uncertainty principle corrections are not tied
down to any specific model of quantum gravity; these corrections
can be derived using arguments from string theory \cite{amati} as
well as other approaches to quantum gravity \cite{magi}.\\
\par\noindent
In recent years, there has been an enormous amount of research
into two important areas of the theoretical physics:branworld
cosmology and the holographic principle. The braneworld scenario
gained momentum as a solution to the hierarchy problem
\cite{{add},{rs1},{rs2}}. The holographic principle, meanwhile,
was first realized in string theory via the AdS/CFT
correspondence\cite{{mal},{wit1},{gab}}. Much attention has
currently been paid for the duality between de Sitter (dS)
gravity and CFT in  analogy to the AdS/CFT correspondence
\cite{ds/cft}(for a very good review see also \cite{odi}). The
essence of braneworld holography {\cite{heb}-\cite{set}} can be
captured in the following claim : Randall-Sundrum branworld
gravity is dual to a CFT with a UV cutoff, coupled to gravity on
the brane. Formal evidence for this claim was provided by
studying a brane universe in the background of the
Schwarzschild-AdS Bblack hole. The introduction of the black hole
on the gravity side of the AdS/CFT correspondence corresponds to
considering finite temperature states in the dual CFT
\cite{wit2}. In the context of braneworld holography, Savonije
and Verlinde demonstrated that their induced braneworld cosmology
could alternatively be described as the standard FRW cosmology
driven by the energy density of this dual CFT
\cite{{verlinde},{savonije}}

In the present paper, we take into account corrections to the
entropy of the  five-dimensional Schwarzschild- anti de Sitter
black hole (abbreviated to $SAdS_5$ in the sequel) that arise due
to the self-gravitational effect. Previous studies of the
Cardy-Verlinde formula (or the corresponding Friedmann equation)
in an AdS/CFT context have attracted a lot of attention. For
instance, the author with Vagenas have shown that the entropy of
Ach\'ucarro-Ortiz black hole (a locally AdS space) can be
described by the Cardy-Verlinde formula \cite{setelias1} and the
self-gravitational corrections to this formula have been computed
\cite{setelias2}. By introducing the self-gravitational
corrections to the Cardy-Verlinde formula, we find a host of
interesting cosmological solutions for the brane universe. Their
most remarkable feature is that the self-gravitational corrections
allows for a nonsingular transition between a contracting phase
of the scale factor of the brane, and a following expanding stage.

\section{Self-Gravitational Corrections to Cardy-Verlind Formula}
In the asymptotic coordinates, the $SAdS_5$ black hole metric is
 \bea
ds^{2}=-F(r)dt^{2}+\frac{1}{F(r)}dr^{2}+r^{2} d\Omega_{(3)}^{2},
\label{metric1} \eea where \be
F(r)=1-\frac{\mu}{r^{2}}+r^{2},\label{fterm} \ee and we work in
units where the $AdS$ radius $l=1$. The parameter $\mu$ is
proportional to the ADM mass $M$ of the black hole. The free
energy, $F$, the entropy, $S$, and the thermodynamical energy
$E$, of the black hole are given respectively, \be
F=\frac{-V_{3}r_{H}^{2}}{16\pi G_5}(r_{H}^{2}-1),\hspace{0.5cm}
S=\frac{V_{3}r_{H}^{3}}{4G_5},\hspace{0.5 cm}E=F+T_{H}S=\frac{3V_3
\mu}{16\pi G_5}.\label{thermo} \ee where $G_5$ is Newton's
constant in $5$ dimensions and $V_3$ is the volume of a unit
$3-$sphere, and  $r_{H}$ is the horizon radius.\\
 We are
interested primarily in the corrections to the entropy
(\ref{thermo}) that arise in the context of KKW analysis
\cite{KKW} \footnote{Further developments in KKW analysis can be
found in \cite{corrections}.} due to self-gravitational effect.
We take into account the response of the background geometry to
an emitted quantum of frequency $\omega$. The particle travels
across the horizon from $r_i$ to $r_f$ which are given by \be
r_i=\sqrt{1/2(\sqrt{1+4\mu}-1) }\label{radi3}  \ee \be
r_f=\sqrt{1/2(\sqrt{1+4\mu'}-1) }.\label{radi4} \ee To first
order in the emitted energy ($\omega$), the afore-mentioned radii
are related as follows \be r^{2}_{f}=r^{2}_{i}\left(1-\omega
\frac{16\pi G_5}{3V_3 r^{2}_{i} \sqrt{1+4\mu}}\right) \ee
Consequently, the change in the entropy of the $SAdS_5$ black hole
during the process of emission takes the form \bea \Delta
S=S_i-S_f=\frac{V_3}{4G_5}(r_{i}^{3}-r_{f}^{3})=\frac{\omega}{T(\omega)}
\label{chang} \eea where $S_f$ is the modified entropy of the
$SAdS_5$ black hole due to the self-gravitational effect, $S_i$ is
the standard formula for the entropy (Bekenstein-Hawking entropy)
derived when the black hole mass is kept fixed, and $T(\omega)$
is the corrected temperature of the black hole horizon $T_{H}$ due
to the self-gravitational effect. The expression for the modified
temperature of the black hole horizon is given as \be
T(\omega)=\frac{\left(2r_{i}^{2}+1\right)}{2\pi r_{i}}+
\mathcal{O}(\omega) \label{modtemp} \ee To first order in
$\omega$, the corrected entropy of the $SAdS_5$ black hole reads
\be S_{f}=S_{i}-\frac{2\pi
r_{i}}{\left(2r_{i}^{2}+l^{2}\right)}\omega \hspace{1ex}.
\label{modentropy} \ee

\par\noindent
Due to the AdS/CFT correspondence, we are now ready to evaluate
the corrections to the Cardy-Verlinde formula for the entropy of
the $SAdS_5$ black hole. The Casimir energy for the spacetime
under study is given by \bea E_C=4E_{4}-3T_{H}S_i \label{cas} \eea
where the four-dimensional energy $E_4$, is given by
\cite{verlinde} \bea E_4=\pm\frac{3V_3}{16\pi G_5
r}\mu=\frac{1}{r}E \label{ener2} \eea where $+$ corresponds to
the black hole horizon and $-$ to the cosmological horizon. It is
easily seen that, to first order in $\omega$, the modified
Casimir energy takes the form \be
\mathcal{E}_{C}=E_{C}-\frac{1}{r}\omega \hspace{1ex}.
\label{modCAS} \ee Due to the self-gravitational corrections, the
modified Cardy-Verlinde formula for the entropy of the $SAdS_{5}$
black hole is given as \be \mathcal{S}_{CFT}=\frac{2\pi
r}{3}\sqrt{{\Big|}\mathcal{E}_C(2\mathcal{E}_4-\mathcal{E}_{C})\Big|}
\hspace{1ex} \label{cv} \ee Using equations (\ref{cas}) and
(\ref{modCAS}), the modified Cardy-Verlinde entropy formula
becomes \be \mathcal{S}_{CFT}=\frac{2\pi r}{3}\sqrt{\Bigg|
\left[E_{C}-\frac{1}{r}\omega\right]
\left[\left(2E_{4}-E_{C}\right)-\frac{1}{r}\omega\right]\Bigg|}
\hspace{1ex} \ee and keeping terms up to first order in the
emitted energy $\omega$, it takes the form

\be \mathcal{S}_{CFT}=S_{CFT}\left(1-\varepsilon \omega\right)
\hspace{1ex} \label{modS} \ee where the small parameter
$\varepsilon$ is given by \be
\varepsilon=\frac{1}{r}\frac{E_{4}}{E_{C}\left(2E_{4}-E_{C}\right)}
\hspace{1ex}. \label{epsilon} \ee A welcomed but not unexpected
result is that there is no entropy bound violation due to
self-gravitational corrections to the Cardy-Verlinde
entropy\footnote{A violation of the holographic entropy bound was
observed when self-gravitational corrections to the
Cardy-Verlinde entropy formula of the two-dimensional
Ach\'ucarro-Ortiz black hole were included \cite{setelias2} (see
also \cite{mignemi}).}.
\section{Self-Gravitational Corrections to FRW Brane Cosmology}
 We now consider a $4$-dimensional brane in the
$SAdS_5$ black hole background. This $4$-dimensional brane can be
regarded as the boundary of the $5$-dimensional $SAdS_{5}$ bulk
background. Let us first replace the radial coordinate $r$ with
$a$ and so the line element (\ref{metric1}) \bea
ds^{2}=-F(a)dt^{2}+\frac{1}{F(a)}da^{2}+a^{2}d\Omega_{(3)}^{2},
\label{met} \eea It was shown that by reduction from the $SAdS_5$
background (\ref{met}) and by imposing the condition \be
-F(a)\left(\frac{\partial t}{\partial \tau}\right)^{2}+
\frac{1}{F(a)}\left(\frac{\partial a}{\partial \tau}\right)^{2}=-1
\label{condition2} \ee where $\tau$ is a new time parameter, one
obtains an FRW metric for the $4$-dimensional timelike brane \be
ds_{(4)}^{2}=-d\tau^{2}+a^{2}(\tau)d\Omega_{(3)}^{2} \hspace{1ex}.
\ee Thus, the $4-$dimensional FRW equation describes the motion
of the brane universe in the $SAdS_5$ background. It is easy to
see that the matter on the brane can be regarded as radiation and
consequently, the field theory on the brane should be a CFT.

\par\noindent
Within the context of context the AdS/CFT correspondence,
Savonije and Verlinde studied the CFT/FRW-cosmology relation from
the Randall-Sundrum type braneworld perspective \cite{savonije}.
They showed that the entropy formulas of the CFT coincides with
the Friedmann equations when the brane crosses the black hole
horizon.

In the case of a $4$-dimensional timelike
 \be
ds_{(4)}^{2}=-d\tau^{2}+a^{2}(\tau)d\Omega_{(3)}^{2} \hspace{1ex},
\ee one of the identifications that supports the
CFT/FRW-cosmology relation  is \be
H^{2}=\left(\frac{2G_4}{V}\right)^{2}\mathcal{S}^{2} \label{huble}
\ee where $H$ is the Hubble parameter defined by
$H=\frac{1}{a}\frac{da}{d\tau}$ and V is the volume of the
$3$-sphere ($V=a^{3}V_{3}$). The $4$-dimensional Newton constant
$G_4$ is related to the $5-$dimensional one $G_5$ by \be
 G_{4}=\frac{2}{l}G_{5}
 \hspace{1ex}.
\ee It was shown that at the moment that the $4$-dimensional
timelike brane crosses the cosmological horizon, i.e. when
$a=a_{b}$, the CFT entropy and the entropy of the $SAdS_5$ black
hole are identical. By substituting (\ref{modS}) into
(\ref{huble}), we obtain the self-gravitational corrections to
the motion of the CFT-dominated brane \be
H^{2}=\left(\frac{2G_4}{V}\right)^{2}S_{CFT}^{2}\left(1-\varepsilon
\omega\right)^{2} \label{modH} \hspace{1ex}. \ee It is obvious
that from the first term on the right-hand side of (\ref{modH})
we get the standard Friedmann equation with the appropriate
normalization \be H^{2}=\frac{-1}{a_{b}^{2}}+\frac{8\pi
G_{4}}{3}\rho \ee where $\rho$ is the energy density defined by
$\rho=E_{4}/V$. Therefore, the correction to the FRW equation due
to the self-gravitation effect is expressed by the second term in
the right-hand side of equation (\ref{modH}). Keeping terms up to
first order in the emitted energy ($\omega$), the modified Hubble
equation due to the self-gravitation corrections is \be
H^{2}=\frac{-1}{a_{b}^{2}}+\frac{8\pi G_{4}}{3}\rho
-2E_{4}\frac{1}{a_{b}}
\left(\frac{2G_{4}}{V}\right)^{2}\left(\frac{2\pi
a_{b}}{3}\right)^{2}\omega \hspace{1ex}. \ee Taking into account
that all quantities should be evaluated on the black hole horizon,
the modified Hubble equation, i.e. the first Friedmann equation,
takes the form \be H^{2}=\frac{-1}{a_{b}^{2}}+\frac{8\pi
G_{4}}{3}\rho - \frac{8\pi G_{4}}{3}\left[\frac{4\pi
G_{4}}{3}\frac{1}{a_{b}^{2}V_{3}}\rho\right]\omega
\label{modHubble} \ee where the volume $V$ is given by
$a_{b}^{3}V_{3}$. At this point it should be stresses that our
analysis was up to now restricted to the spatially flat ($k=+1$)
spacelike brane.
\par
We will now extend the aforesaid analysis. We therefore consider
an arbitrary scale factor $a$ and include a general $k$ taking
values ${+1,0,-1}$ in order to describe, respectively, the
elliptic, flat, and hyperbolic horizon geometry of the $SAdS_5$
bulk black hole. The modified Hubble equation is now given by \be
H^{2}=\frac{-k}{a^{2}}+\frac{8\pi G_{4}}{3}\rho - \frac{8\pi
G_{4}}{3}\left[\frac{4\pi
G_{4}}{3}\frac{1}{a^{2}V_{3}}\rho\right]\omega \label{gmodHubble}
\ee where the volume $V$ is now given by $a^{3}V_{3}$   since all
quantities that appear in equation (\ref{gmodHubble}) are defined
for an arbitrary scale factor $a$.

The first term in the right-hand side of equation
(\ref{gmodHubble}) represents the curvature contribution to the
brane motion. The second term can be regarded as the contribution
from the radiation and  it redshifts as $a^{-4}$. The last term in
the right-hand side of equation (\ref{gmodHubble}) is the
self-gravitational correction to the motion of $4$-dimensional
timelike brane moving in the $5$-dimensional Schwarzschild-anti de
Sitter bulk background. Since this term goes like $a^{-6}$, it is
obvious that it is dominant at early times of the brane evolution
while at late times the second term, i.e. the radiative matter
term, dominates and thus the last term can be neglected. The sign
of last term is opposite with respect to the standard situation,
one may expect that this sign difference could have interesting
cosmological consequences. The bounce can be attributed to the
negative-energy matter, which dominates at small values of $a$ and
create a significant enough repulsive force so that a big crunch
is avoided. Indeed, we will see that it is crucial in allowing a
nonsingular transition between a contracting and an expanding
evolution of the scale factor $a$.

\par\noindent
We would like  to regard the last term in right-hand side of
equation (\ref{gmodHubble}) as stiff matter \cite{stiff1}. In
particular, within the context of AdS/CFT
correspondence,Mukherji, Peloso \cite{mukh} and Medved\cite{medve}
have considered a Reissner-Nordstrom-anti de Sitter bulk
background \cite{stiff2} \be
ds^{2}=-h(a)dt^{2}+\frac{1}{h(a)}da^{2}+a^{2}d\Omega_{(3)}^{2},
\ee where \be
h(a)=k+\frac{a^{2}}{l^{2}}-\frac{\varepsilon_{3}M}{a^{2}}+\frac{3\varepsilon_{3}^{2}Q^{2}}{16a^{4}}
\hspace{1ex}. \ee where \be \varepsilon_{3}=\frac{16\pi G_5}{3V_3}
\ee and $M$, $Q$ represent the black hole mass and charge,
respectively.\\
For charged black hole bulk, the brane world turns out to be a
bounce cosmology \cite{{bounce},{mukh}, {medve}}. That is, the
universe is asymptotically de Sitter in the far past, contracts
to a non-vanishing minimum at some given time and then expands to
an asymptotically de Sitter future. It was showed that the brane
evolution is described by a Friedmann-like equation for radiative
matter along with a stiff-matter contribution \bea
H^{2}&=&\frac{-k}{a^{2}}+\frac{\varepsilon_{3}M}{a^{4}}-
\frac{3\varepsilon_{3}^{2}Q^{2}}{16a^{6}}\nn\\
&=&\frac{-k}{a^{2}}+\frac{8\pi G_{4}}{3}\rho-
\frac{3\varepsilon_{3}^{2}Q^{2}}{16a^{6}} \label{stiff}
\hspace{1ex}. \eea If the following condition is satisfied \be
Q^{2}=\frac{8}{3}M\omega \hspace{1ex}, \ee then equations
(\ref{gmodHubble}) and (\ref{stiff}) are identical and the last
term in right-hand side of equation (\ref{gmodHubble}) is then
regarded as stiff matter.\\
The evolution of the system can be solved exactly, as one can
most simply realize by using conformal time $\eta$, defined as
$dt=a(\eta)d\eta$. Let us first consider a closed four
dimensional geometry. In this case the solution is \be
a(\eta)=\sqrt{\frac{\varepsilon_{3} M}{2}}[1-C_1
\cos(2\eta)]^{1/2}, \hspace{1cm}C_1=\sqrt{1-\frac{2\omega}{M}}.
\label{sol1} \ee  $C_1$ is real because $2\omega \ll M$. Hence,
the universe evolves periodically, with a four dimensional radius
oscillating between a maximal and a minimal size given by \be
a_{max,min}=\sqrt{\frac{\varepsilon_{3} M}{2}} (1\pm
C_1)^{1/2}.\label{maxmin} \ee Notice that we have used the
freedom in setting the origin of conformal time so that the
minimal radius is reached at $\eta=n\pi$, where $n$ is an integer
number. From the bulk perspective, the brane starts out from the
black hole at a distance $a_{min}$ from the singularity and moves
away up to $a_{max}$ as it expands. At later time, it collapses
at $a=a_{min}$. It is interesting to compare the case at hand and
the situation without self-gravitational corrections. In the
latter situation, the cosmological evolution is governed by
Eq.(\ref{gmodHubble}), but with $\omega=0$. This, in turn, means
that the time dependence of the scale factor is given by
Eq.(\ref{sol1}), with $C_{1}=1$. In this case, the scale factor
starts from zero size and expands up to $a_{max}$ before
collapsing again to zero size. From the bulk point of view, the
brane originates from the black hole singularity and at a latter
stage it collapses again into the singularity. We, therefore, see
that the effect of self-gravitational corrections is rather
non-trivial, since it
make the cosmology of the model free from singularities.\\
For an open universe, we find the solution \be
a(\eta)=\sqrt{\frac{\varepsilon_{3} M}{2}}[C_2
\cosh(2\eta)-1]^{1/2}, \hspace{1cm}C_2=\sqrt{1+\frac{2\omega}{M}}.
\label{sol2} \ee In this case, the brane is initially
contracting, and then bounces to an expanding phase. Again, we
have set $\eta=0$ at the bounce. The minimal radius is given by
\be a_{min}=\sqrt{\frac{\varepsilon_{3} M}{2}}(C_2-1)^{1/2}
\label{min} \ee As before, $a_{min}\rightarrow 0$ as
$\omega\rightarrow 0$.\\
Finally, in the case of a flat universe we find \be
a(\eta)=\sqrt{\varepsilon_{3}(M\eta^{2}+\frac{\omega}{2})}
\label{sol3} \ee Also in this case we have a bouncing universe,
with a minimal radius \be a_{min}=\sqrt{\frac{\varepsilon_{3}
}{2}\omega} \label{minf}\ee At late times one recovers the
evolution $a(\eta)\sim \eta\sim t^{1/2}$, which is typical of a
flat universe dominated by radiation.\\
It is clear from the above forms that, as long as the
self-gravitational corrections are non-vanishing, $a$ will never
shrink to zero. At the bounce ($t=\eta=0$) in particular, all
three equations take the simple form \be
a(0)=\sqrt{\frac{\varepsilon_{3}}{2}\omega}+O(\omega^{2}).
\label{simpl}\ee Therefore, the feature of a bouncing universe
would allow one to circumvent the issue of resolving the big bang
(or crunch) singularity, which afflicts many (if not most)
cosmological models.\\

\section{Conclusions}
One of the striking results for the dynamic AdS/CFT correspondence
is that the Cardy-Verlinde's formula on the CFT-side coincides
with the first Friedmann equation (Hubble equation) in cosmology
when the brane crosses the horizon $a=a_{b}$ of the $SAdS_5$ black
hole. This means that the Hubble equation knows the
thermodynamics of the CFT. In this paper we have considered the
dynamics of a $4$-dimensioanl timelike FRW brane propagating in an
$5$-dimensional AdS  bulk space containing  a Schwarschild black
hole. Taking into account the semiclassical corrections to the
black hole entropy that arise as a result of the
self-gravitational effect,  and employing the AdS/CFT
correspondence, we obtained the self-gravitational corrections to
the Cardy-Verlinde formula. A welcomed but not unexpected result
was that the modified entropy doesn't violate any entropy bound
since the additional term due to the self-gravitation effect is
subtractive. These self-gravitational corrections to the
Cardy-Verlinde entropy formula express the existence of a deep
connection between semi-classical thermodynamics and Anti de
Sitter holography.

\par\noindent
Furthermore, the self-gravitational corrections to the associated
Friedmann-like brane equations are obtained. The additional term
in the Hubble equation due to the self-gravitation effect goes as
$a^{-6}$. Thus, the self-gravitational corrections act as a
source of stiff matter contrary to standard FRW cosmology where
the charge of the black hole plays this role. A very interesting
feature of the solutions of FRW equation is that they all (for
$k=0,\pm1$) have a non-vanishing minimum value for the scale
factor. That is to say, as long as the self-gravitational
correction is non-vanishing, a singularity will certainly be
avoided. As a consequence, we found the cosmological evolution of
the universe to be regular, with a smooth transition between a
contracting and an eventual expanding phase. We presented exact
cosmological solutions for a (open, flat, and closed) critical
brane.



\begin{thebibliography}{99}
\bibitem{maj1x} R.K.  Kaul and P. Majumdar, Phys. Rev. Lett. {\bf 84}, 5255
(2000).
\bibitem{carx} S. Carlip, Class. Quant. Grav. {\bf 17}, 4175
(2000).
\bibitem{das2x} S. Das, P. Majumdar and R.K. Bhaduri, Class. Quant. Grav.
{\bf 19}, 2355 (2002).
\bibitem{das1x}
 S. Das, R.K. Kaul and P. Majumdar, Phys. Rev. {\bf D63}, 044019 (2001).
\bibitem{kun2x} A.J.M. Medved and G. Kunstatter, Phys. Rev. {\bf D63},
104005 (2001).
\bibitem{bir2x} D. Birmingham, I. Sachs and S. Sen,
Int. J. Mod. Phys. {\bf D10}, 833 (2001).
\bibitem{govx}
T.R. Govindarajan, R.K. Kaul and V. Suneeta, Class. Quant. Grav.
{\bf 18}, 2877 (2001).
\bibitem{cavx} M. Cavaglia and A. Fabbri, Phys. Rev. {\bf D65},
044012 (2002).
\bibitem{ajm1x} A.J.M. Medved, Class. Quant. Grav. {\bf 19}, 2503
(2002).
\bibitem{mukx} S. Mukherji and S. S. Pal, JHEP {\bf 0205}, 026
(2002).
\bibitem{das3x} S. Das, hep-th/0207072.
\bibitem{gg2x} G. Gour, Phys. Rev. {\bf D66}, 104022 (2002).
\bibitem{maj3x} A. Chatterjee and P. Majumdar, gr-qc/0303030.
\bibitem{od}J. E. Lidsey, S. Nojiri, S. D. Odintsov and S. Ogushi,
Phys. Lett. {\bf B544}, 337, (2002).
\bibitem{medved}G. Gour, A. J. M. Medved, gr-qc/0305018.
\bibitem{od1} S. Nojiri, S. D. Odintsov, S. Ogushi,
hep-th/0212047.
\bibitem{set5}M. R. Setare, Phys. Lett. {\bf B573}, 173 (2003).
\bibitem{set6}M. R. Setare, Eur. Phys. J. {\bf C33}, 555, (2004).
\bibitem{carli}S. Carlip, Class. Quant. Grav. {\bf 15}, 3609,
(1998).
\bibitem{hawking1}
S.W. Hawking, Commun. Math. Phys. {\bf43} (1975) 199.


\bibitem{KKW}
P. Kraus and F. Wilczek, Nucl. Phys. B {\bf433} (1995) 403,
gr-qc/9408003; P. Kraus and F. Wilczek, Nucl. Phys. B {\bf437}
(1995) 231, hep-th/9411219; E. Keski-Vakkuri and P. Kraus, Nucl.
Phys. B {\bf491} (1997) 249, hep-th/9610045; M.K. Parikh and F.
Wilczek, Phys. Rev. Lett. {\bf85} (2000) 5042, hep-th/9907001.

\bibitem{corrections}
Y. Kwon, Il Nuovo Cimento B {\bf115} (2000) 469; S. Hemming and E.
Keski-Vakkuri, Phys. Rev. D {\bf64} (2001) 044006, gr-qc/0005115;
E.C. Vagenas, Phys. Lett. B {\bf503} (2001) 399, hep-th/0012134;
E.C. Vagenas, Phys. Lett. B {\bf533} (2002) 302, hep-th/0109108;
A.J. Medved, Class. Quant. Grav. {\bf19} (2002) 589,
hep-th/0110289; E.C. Vagenas, Phys. Lett. B {\bf559} (2003) 65,
hep-th/0209185; M.K. Parikh, hep-th/0402166; M. R. Setare and E.
C. Vagenas, hep-th/0405186.
\bibitem{amati}D. Amati, M. Ciafaloni and G. Veneziano, Phys.
Lett. {\bf B216}, 41 (1989); D. Amati, M. Ciafaloni and G.
Veneziano, Nucl. Phys. {\bf B347}, 550 (1990), K. Konishi, G.
Paffuti and P. Provero, Phys. Lett. {\bf B234}, 276 (1990).
\bibitem{magi} M. Maggiore, Phys. Rev. {\bf D49}, 5182 (1994); M. Maggiore,
Phys. Lett. {\bf B319}, 83 (1993).
\bibitem{add} N. Arkani-Hamed, S. Dimopoulos, G. Dvali, Phys. Lett. {\bf B429}, 263, (1998).
\bibitem{rs1}L. Randall, R. Sundrum, Phys. Rev. Lett. {\bf 83}, 3370, (1999).
\bibitem{rs2}L. Randall, R. Sundrum, Phys. Rev. Lett. {\bf 83}, 4690, (1999).
\bibitem{mal}J. M. Maldacena, Adv. Theor. Math. Phys. {\bf 2}, 231, (1998).
\bibitem{wit1}E. Witten, Adv. Theor. Math. Phys. {\bf 2}, 253, (1998).
\bibitem{gab} S. S. Gubser, I. R. Klebanov, A. M. Polyakov, Phys. Lett. {\bf B428}, 105, (1998).

\bibitem{ds/cft}
A. Strominger, JHEP {\bf0110} (2001) 034, hep-th/0106113; A.
Strominger, JHEP {\bf0111} (2001) 049, hep-th/0110087; For review,
see M. Spradlin, A. Strominger and A. Volovich, hep-th/0110007; S.
Nojiri and S.D. Odintsov, Phys. Lett.B {\bf519} (2001) 145,
hep-th/0106191; S. Nojiri and S.D. Odintsov, JHEP {\bf0112} (2001)
033, hep-th/0107134; S. Nojiri, S.D. Odintsov and S. Ogushi, Phys.
Rev. D {\bf65} (2002) 023521, hep-th/0108172; S. Nojiri and S.D.
Odintsov, Phys. Lett. B {\bf523} (2001) 165, hep-th/0110064; D.
Klemm, Nucl. Phys. B {\bf625} (2002) 295, hep-th/0106247; T.
Shiromizu, D. Ida and T. Torii, JHEP {\bf0111} (2001) 010,
hep-th/0109057; C.M. Hull, JHEP {\bf0111} (2001) 012,
hep-th/0109213; B. McInnes, Nucl. Phys. B {\bf627} (2002) 311,
hep-th/0110062; Y. S. Myung, Mod. Phys. Lett. A {\bf16} (2001)
2353, hep-th/0110123; E. Halyo, JHEP {\bf0203} (2002) 009,
hep-th/0112093; C. P. Burgess, F. Quevedo, R. Rabadan, G. Tasinato
and I. Zavala, JCAP,  {\bf0402}, 008, (2004);  O. Obregón, L.
Patiño and H. Quevedo. Phys. Rev. {\bf D68}, 026002, (2003); M. R.
Setare, Mod. Phys. Lett. A {\bf17} (2002) 2089, hep-th/0210187; M.
R. Setare and M. B. Altaie, Eur. Phys. J. C {\bf30} (2003) 273,
hep-th/0304072; M. R. Setare, hep-th/0405010.

\bibitem{odi}
S. Nojiri, S.D. Odintsov and S. Ogushi, Int. J. Mod. Phys. A
{\bf17} (2002) 4809, hep-th/0205187.
\bibitem{heb} A. Hebecker, J. March-Russell, Nucl. Phys. {\bf B608}, 375, (2001).
\bibitem{savonije}
I. Savonije and E. Verlinde, Phys. Lett. B {\bf507} (2001) 305,
hep-th/0102042.
\bibitem{pad1}A. Padilla, Phys. Lett. {\bf B528}, 274, (2002).
\bibitem{pad1}A. Padilla, hep-th/0210217.
\bibitem{set}M. R. Setare and R.
Mansouri, Int. J. Mod. Phys.  A {\bf18} (2003) 4443,
hep-th/0210252.
\bibitem{wit2}E. Witten, Adv. Theor. Math. Phys. {\bf 2}, 505, (1998).
\bibitem{verlinde} E. Verlinde, hep-th/0008140.


\bibitem{setelias1}
M. R. Setare and E. C. Vagenas, Phys.Rev.D {\bf68} (2003) 064014,
hep-th/0304060.

\bibitem{setelias2}
M. R. Setare and E. C. Vagenas Phys. Lett. {\bf B584} (2004) 127,
hep-th/0309092.


\bibitem{BBM}
V. Balasubramanian, J. de Boer and D. Minic, Phys. Rev. D {\bf65}
(2002) 123508, hep-th/0110108.




\bibitem{medved1} A.J.M.  Medved, Phys. Rev. D {\bf66} (2002) 124009, hep-th/0207247.


\bibitem{cvetic}
M. Cveti$\check{c}$, S. Nojiri and S.D. Odintsov, Nucl. Phys. B
{\bf628} (2002) 295, hep-th/0112045; S. Nojiri and S.D. Odintsov,
gr-qc/0112066.


\bibitem{mignemi} S.~Mignemi, to appear in Phys. Rev. D, hep-th/0307205.

\bibitem{mukh}S. Mukherji and M. Peloso, Phys. Lett. B {\bf547} (2002) 297,
hep-th/0205180.
\bibitem{medve}A.J.M.  Medved, JHEP, O305, 008, (2003),
hep-th/0301010.
\bibitem{bounce}A.J.M.  Medved, hep-th/0205251;
Y. Shtanov and V. Sahni, Phys. Lett. B {\bf557} (2003) 1,
gr-qc/0208047.


\bibitem{stiff1} A.K. Biswas and S. Mukherji, JHEP {\bf0103} (2001)
046, hep-th/0102138.

\bibitem{stiff2}
A.J.M.  Medved, hep-th/0111182.



\end{thebibliography}
\end{document}